\documentclass[10pt,a4paper]{article}

\usepackage[latin1]{inputenc}
\usepackage{fourier}
\usepackage{graphicx}
\usepackage{amsmath}
\usepackage{amsfonts}
\usepackage[latin1]{inputenc}
\usepackage[T1]{fontenc}
\usepackage{verbatim}
\usepackage{latexsym}
\usepackage{natbib}
\usepackage{bm}
\usepackage[labelfont=bf]{caption}
\usepackage{multirow}
\usepackage{array}
\usepackage{enumitem}

\usepackage{hyperref}
\hypersetup{backref,colorlinks=true,citecolor=blue}

\usepackage{alltt}
\usepackage{color}                       
\definecolor{grey}{rgb}{0.9,0.9,0.9}

\usepackage{etoolbox}
\makeatletter
\patchcmd{\NAT@citex}
  {\@citea\NAT@hyper@{%
     \NAT@nmfmt{\NAT@nm}%
     \hyper@natlinkbreak{\NAT@aysep\NAT@spacechar}{\@citeb\@extra@b@citeb}%
     \NAT@date}}
  {\@citea\NAT@nmfmt{\NAT@nm}%
   \NAT@aysep\NAT@spacechar\NAT@hyper@{\NAT@date}}{}{}
\patchcmd{\NAT@citex}
  {\@citea\NAT@hyper@{%
     \NAT@nmfmt{\NAT@nm}%
     \hyper@natlinkbreak{\NAT@spacechar\NAT@@open\if*#1*\else#1\NAT@spacechar\fi}%
       {\@citeb\@extra@b@citeb}%
     \NAT@date}}
  {\@citea\NAT@nmfmt{\NAT@nm}%
   \NAT@spacechar\NAT@@open\if*#1*\else#1\NAT@spacechar\fi\NAT@hyper@{\NAT@date}}
  {}{}
\makeatother

\begin{document}

\title{The CGEM Lorentz Force Data from HMI Vector Magnetograms}

\author{Xudong Sun$^{a,b}$ for the CGEM Team \\
\small $^{a}$ HEPL, Stanford University, Stanford, CA 94305 \\
\small $^{b}$ Now at: IfA, University of Hawaii, Pukalani, HI 96768 (\href{mailto:xudongs@hawaii.edu}{xudong@hawaii.edu})}

\date{\normalsize \textit{June 9, 2022}}

\maketitle

\begin{abstract}
We describe a new data product from the CGEM (Coronal Global Evolutionary Model) collaboration that estimates the Lorentz force in active regions (ARs) based on HMI vector magnetogram patches. Following Fisher et al. (2012), we compute three components of the integrated Lorentz force over the outer solar atmosphere every 12 minutes throughout an AR's disk passage. These estimates, differenced during solar eruptive events, can provide valuable diagnostics on dynamic processes. We describe the pipeline modules, provide data retrieval examples, and document some systematic uncertainties that users should be aware of. Finally we document the formal uncertainty propagation procedures.
\end{abstract}

\section{Lorentz Force on Plasma Bulk}
\label{sec:force}

Starting from the Maxwell stress tensor for a static magnetic field, the total Lorentz force acting on a bulk of plasma in the solar atmosphere (at and above the photosphere) can be written as a surface integral of the photosphere field $\bm{B}=(B_x,B_y,B_z)$, assuming the contribution from the top and side boundaries is negligible \citep{fisher2012}. The horizontal ($F_x$, $F_y$) and vertical ($F_z$) components are
\begin{equation}
\label{eq:lorentz}
\begin{split}
F_x & = - \frac{1}{4\pi} \int_A B_x B_z \, \mathrm{d}A, \\
F_y & = - \frac{1}{4\pi} \int_A B_y B_z \, \mathrm{d}A, \\
F_z & = \frac{1}{8\pi} \int_A (B_x^2 + B_y^2 - B_z^2) \, \mathrm{d}A,
\end{split}
\end{equation}
respectively. Here the $z$-direction is pointing upward from the photosphere, and $A$ is the pixel area as weight. For magnetograms in the observation (CCD image) plane, there is $A \propto \mu^{-1}$, where $\mu$ is cosine of the angle between line-of-sight (LOS) and the local normal. For magnetograms with equal-area pixels (e.g., cylindrical equal-area projection), $A$ can be dropped.

Recent observations have confirmed that rapid, permanent changes take place in the photospheric field during major flares \citep{sudol2005,petrie2010,wangs2012,sun2012}. This change results in a Lorentz force impulse \citep{hudson2008,fisher2012}; the change of force $\delta \bm{F} = (\delta F_x, \delta F_y, \delta F_z)$ is
\begin{equation}
\label{eq:df}
\begin{split}
\delta F_x & = - \frac{1}{4\pi} \int_A \delta (B_x B_z) \, \mathrm{d}A, \\
\delta F_y & = - \frac{1}{4\pi} \int_A \delta (B_y B_z) \, \mathrm{d}A, \\
\delta F_z & = \frac{1}{8\pi} \int_A (\delta B_x^2 + \delta B_y^2 - \delta B_z^2) \, \mathrm{d}A.
\end{split}
\end{equation}

During major flares, $B_h=\sqrtsign{B_x^2+B_y^2}$ generally increases, while $B_z$ remains less effected. This results in a positive $\delta F_z$, the impulse of which may help launch the ejecta. One may estimate the CME mass $M$, for example, from
\begin{equation}
\label{eq:mcme}
M \sim \frac{\delta F_z \delta t}{2 v},
\end{equation}
where $\delta t$ is the interaction time scale (e.g. minutes), and $v$ the CME speed. As another example, the change of torque from $\bm{F}_h$ has been used to explain the sudden angular velocity change of the rotating sunspots during a large flare \citep{wangs2014}.

It is worthwhile to note that the Lorentz force or Lorentz force impulse acting on the solar interior (at the photosphere and below) are equal and opposite to the values given by Equation set~\ref{eq:lorentz}.



\begin{figure}[th]
\centerline{\includegraphics{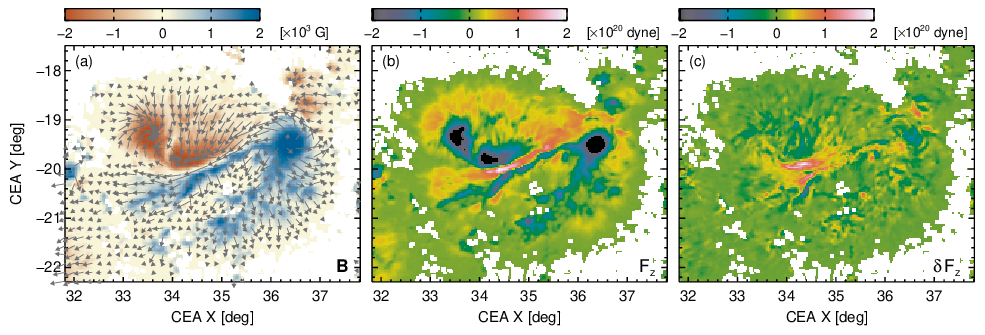}}
\caption{Center portion of the CEA maps from SHARP 377 (NOAA AR 11158) on 2011 Feb 15. Weak field pixels are masked out and shown as white. (a) Vector field map for 01:36 UT, right before a X-class flare. (b) Vertical Lorentz force for 01:36 UT. Values at each pixel are calculated as $(B_\phi^2 + B_\theta^2 - B_r^2) \Delta A$, where $\Delta A$ is the pixel area. At the default 0.03$^\circ$, it's about $1.3 \times 10^{15}$~cm$^2$. (c) Difference map between 02:36 UT (post-flare) and 01:36 UT (pre-flare). Large increase is seen near the central polarity inversion line. \label{f:map}}
\end{figure}


\section{The CGEM Data Product: \texttt{cgem.Lorentz}}
\label{sec:data}

The HMI pipeline \citep{hoeksema2014} now routinely produces full disk, disambiguated vector magnetograms. We make use of the Space weather HMI Active Region Patch \citep[SHARP;][]{bobra2014} data product, which automatically extracts and groups the active region vector magnetogram patches. We perform mapping and vector projection on the SHARP data \citep{sun2013}; the final product includes three maps $(B_\phi, B_\theta, B_r)$ (zonal, meridional, radial) measured on a cylindrical equal-area (CEA) grid. We make the approximation $(B_\phi, B_\theta, B_r)=(B_x,-B_y,B_z)$ and use these vector maps to estimate the Lorentz force based on Equation set~\ref{eq:lorentz} for each AR every 12 minutes.

The data series \href{http://jsoc.stanford.edu/ajax/lookdata.html?ds=cgem.lorentz}{\texttt{cgem.Lorentz}} consists of the following, and can be accessed via the \href{http://jsoc.stanford.edu}{JSOC website}.
\begin{enumerate}[noitemsep,topsep=2pt,parsep=1pt,partopsep=1pt,label=\textit{\alph*}),leftmargin=*]
\item Integrated Lorentz forces as keywords. Table~\ref{tbl:keys} lists the relevant keywords with descriptions. Two sets of calculations are involved: one uses every pixel in the map; the other uses only the strong field pixels based on a time-space-dependent noise mask\footnotemark. A normalized version is also provided, which is useful for evaluating the magnetoram as input for force-free extrapolations \citep{wiegelmann2006}.
\item Three "force density" maps (e.g. a map of $-B_x B_z \Delta A / (4 \pi)$). Figure~\ref{f:map} shows an example.
\end{enumerate}

\footnotetext{The noise mask is generated for the 180-degree azimuthal ambiguity resolution and is described in Section 7.1.1 in \citet{hoeksema2014}. In short, we gather field strength images taken where the spacecraft is with in a certain velocity range, smooth them with low order Chebychev filter, and use the median value of all these images at each pixel as the noise estimation. They are named \texttt{conf\_disambig} and are available in the SHARP data series \href{http://jsoc.stanford.edu/ajax/lookdata.html?ds=hmi.sharp_cea_720s}{\texttt{hmi.sharp\_cea\_720s}}. A value of 90 indicates strong field.}


\begin{table}[th]
\begin{center}
\caption{Integrated Lorentz force as keyword, with $(B_\phi, B_\theta, B_r)=(B_x,-B_y,B_z)$ \label{tbl:keys}}
\renewcommand{\arraystretch}{1.2}
\begin{tabular}{l|l|c|l}
\hline
\hline
All pix & Strong pix & Unit & Value \\
\hline
\texttt{TOTFX} & \texttt{TOTFX1} & \multirow{3}{*}{$10^{20}$ dyne} & -$\sum B_\phi B_r \Delta A / (4 \pi)$ \\
\texttt{TOTFY} & \texttt{TOTFY1} & & $\sum B_\theta B_r \Delta A / (4 \pi)$ \\
\texttt{TOTFZ} & \texttt{TOTFZ1} & & $\sum (B_\phi^2 + B_\theta^2 - B_r^2) \Delta A / (8 \pi)$ \\
\hline
\texttt{TOTBSQ} & \texttt{TOTBSQ1} & G$^2$  & $\sum B^2$ \\
\hline
\texttt{EPSX} & \texttt{EPSX1} & \multirow{3}{*}{none} & $\sum B_\phi B_r / \sum B^2$ \\
\texttt{EPSY} & \texttt{EPSY1} & & -$\sum B_\theta B_r / \sum B^2$ \\
\texttt{EPSZ} & \texttt{EPSZ1} & & $\sum (B_\phi^2 + B_\theta^2 - B_r^2) / \sum {B^2}$ \\
\hline
\end{tabular}
\end{center}
\end{table}

\begin{figure}[th]
\centerline{\includegraphics{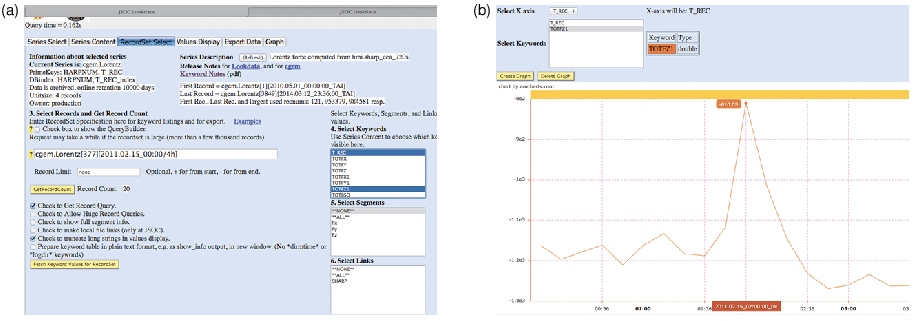}}
\caption{Screen captures from the JSOC website. (a) ``RecordSet Select'' tab. Keywords \texttt{T\_REC} and \texttt{TOTFZ1} are highlighted. Twenty records for SHARP number 377 are selected with query cgem.Lorentz[377][2011.02.15\_00:00/4h]. One may also use the following template to query on NOAA numbers: cgem.Lorentz[][2011.02.15\_00:00/4h][?  NOAA\_ARs like "11158" ?].  (b) ``Graph'' tab. Four-hour profile of \texttt{TOTFZ1} is generated, where an increase of $F_z$ is seen. The value of $F_z$ peaks around 02:00 UT and returns to the pre-flare value by 02:36 UT. \label{f:jsoc}}
\end{figure}

The quick-plot service provided by the JSOC website is useful for exploring the temporal evolution. Users may select the keywords they wish to explore under the ``RecordSet Select'' tab, and use the ``Graph'' tab to generate the temporal profile. Figure~\ref{f:jsoc} shows an example.


\begin{figure}[th]
\vspace*{0.5cm}
\centerline{\includegraphics{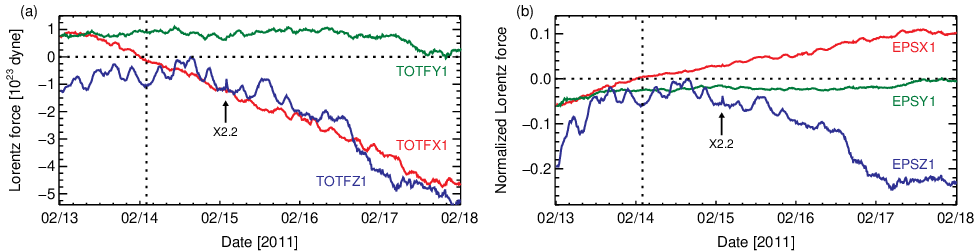}}
\caption{Evolution of integrated Lorentz forces for SHARP 377 (NOAA AR 11158). (a) Lorentz forces for strong field. The evolution is dominated by the secular variations presumably from noise level changes and spacecraft orbital velocity. Time of central meridian passage is marked by vertical dotted line, and the X-class flare is marked with an arrow. (b) Normalized Lorentz forces for strong field. Note that \texttt{EPSX1} and \texttt{EPSY1} are defined such that their signs are opposite those of \texttt{TOTALFX1} and \texttt{TOTALFY1} (Table~\ref{tbl:keys}). \label{f:secular}}
\end{figure}


\section{Usage and Uncertainties}
\label{sec:usage}

Several systematic uncertainties are listed here. An example temporal profile illustrates these points in Figure~\ref{f:secular}.

\begin{enumerate}[noitemsep,topsep=0pt,parsep=4pt,partopsep=4pt,leftmargin=*]

\item \textit{Orbital-velocity-related periodicity}. Due to the geosynchronous orbit of SDO, the line profile shifts throughout the day. This affects the magnetic field determination, mostly in the line-of-sight component \citep{hoeksema2014}. The variation is estimated to be about 1$\%$ in the umbrae, 2$\%$ in the penumbrae, and 5$\%$ in the quiet regions. Such variation is carried into the Lorentz force calculation.

\item \textit{Longitudinal dependence of $F_x$ and $F_z$}. We find that the integrated $F_x$ is generally positive on the eastern hemisphere, and negative on the west. The values scale approximately linearly with longitude, and the relative magnitude $\sum (B_x B_z) / \sum B^2$ (\texttt{EPSX}) can be several tens of percent when the AR is away from the central meridian\footnotemark. This signal thus dominates other known systematics. $F_z$ is also affected but not so much for $F_y$. We tentatively attribute this to the changing noise level in $B_x$. Investigation is underway.

\item \textit{Mapping and masking}. The calculations are performed in CEA coordinates \citep{sun2013}. An independent Mercator mapping was tested which yields differences of a few percent. During the transition from AR patch to full-disk azimuthal disambiguation, the weak-field mask threshold has increased. (The AR patch disambiguation uses the noise mask plus a 20~G constant as the threshold, whereas for full disk the constant increases to 50~G to speed up the computation.) This yields a difference of several percent, and affects data after 2014 Jan 15.

\end{enumerate}

\footnotetext{To rule out the effect of field evolution, we have studied about 20 mature sunspots. We found that the trend persists. This can be a problem when the data are used as input for magnetic extrapolation models.}


\begin{figure}[th]
\centerline{\includegraphics{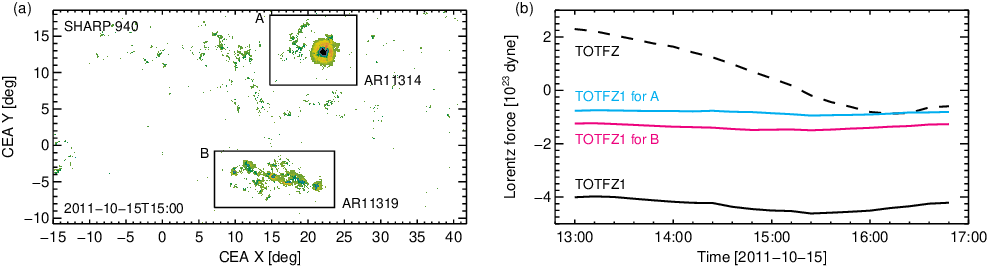}}
\caption{Example of a SHARP including multiple NOAA ARs and large quiet regions. (a) Map of $F_z$ for SHARP number 940, including NOAA AR 11314 and 11319. White color indicates area that is masked as weak field. The two subregions A and B are marked by two boxes. (b) Four-hour evolutions of \texttt{TOTFZ} and \texttt{TOTFZ1} for the whole region, and \texttt{TOTFZ1} for subregions A and B. \label{f:multi}}
\end{figure}


A few further notes.

\begin{enumerate}[noitemsep,topsep=0pt,parsep=4pt,partopsep=4pt,leftmargin=*]

\item The SHARP maps may consist of multiple ARs with large quiet regions in between. The integrals from Equation set~\ref{eq:lorentz} may be very different if these quiet regions are included (e.g. \texttt{TOTFZ} vs. \texttt{TOTFZ1}), as illustrated in Figure~\ref{f:multi}. It can also be very different for individual subregions. Conclusions from the indices should be checked against the maps. 

\item The sudden changes of $F_z$ (over about one hour) can be small compared to the secular ones (over days). Results need to be interpreted with care.

\item Non-optimal data (flagged with a non-zero value of the QUALITY keyword) may have higher noise and thus abnormal values of Lorentz force.

\item The increase of $F_z$ near the polarity inversion line generally persists, as the increase of $B_h$ there is permanent. However, the integrated $F_z$ may display a spike-like profile, returning to the pre-flare values soon after the eruption (Figure~\ref{f:jsoc}(b)). Difference maps show that the increase near the polarity inversion line is compensated by the decrease in peripheral regions (Figure~\ref{f:map}(c)).

\item The increasing noise level in the vector field towards limb and the varying noise-mask yield a variation of the number of pixels included in the computation. Care is needed for interpreting data far away from central meridian.

\item The CEA maps retain the same map size for all time steps, resulting frames containing \texttt{NaN}s near the limb. In this case the indices may be \texttt{NaN} too.

\end{enumerate}


\section{Formal Uncertainty Estimates\protect\footnotemark}
\label{sec:errors}

\footnotetext{Added on Oct 9, 2018. Not implemented in the pipeline.}

Another useful estimation is the relative Lorentz force, $\bm{\epsilon} = (\epsilon_x, \epsilon_y, \epsilon_z)$, which is useful for evaluating the magnetogram as input for force-free extrapolations \citep{wiegelmann2006}:
\begin{equation}
\label{eq:epsilon}
\begin{split}
\epsilon_x & = \frac{ \sum B_x B_z A } {\sum (B_x^2 + B_y^2 + B_z^2) A}, \\
\epsilon_y & = \frac{ \sum B_y B_z A } {\sum (B_x^2 + B_y^2 + B_z^2) A}, \\
\epsilon_z & = \frac{ \sum (B_x^2 + B_y^2 - B_z^2) A } {\sum (B_x^2 + B_y^2 + B_z^2) A}, \\
\epsilon & = |\epsilon_x| + |\epsilon_y| + |\epsilon_z|.
\end{split}
\end{equation}
There is obviously $-0.5 \leq \epsilon_x, \epsilon_y \leq 0.5$, $-1 \leq \epsilon_z \leq 1$ and $0 \leq \epsilon \leq 1$.

The HMI vector magnetograms are derived from spectral inversion and removed of the 180$^\circ$ azimuthal ambiguity. The three components of the field vectors are provided in field strength $B$, inclination $\gamma$, and azimuth $\psi$. Errors ($\sigma_B$, $\sigma_\gamma$, $\sigma_\psi$) and correlation coefficients ($\rho_{B\gamma}$, $\rho_{B\psi}$, $\rho_{\psi\gamma}$) from the inversion are provided as uncertainty estimation \cite{centeno2014}.

Equation sets (1), (7), and (8) in \citet{sun2013} give the links between $(B, \gamma, \psi)$ and $(B_x,B_y,B_z)$. We repeat those below. First, we get the three Cartesian components in the observational plane $(B_\xi, B_\eta, B_\zeta)$:
\begin{equation}
\label{eq:bvec}
\begin{split}
B_\xi & = - B \sin\gamma \sin \psi, \\
B_\eta & = B \sin\gamma \cos \psi, \\
B_\zeta & = B \cos\gamma.
\end{split}
\end{equation}
Then, the field vectors are transformed into the Heliocentric spherical coordinate $(B_\phi,B_\theta,B_r)=(B_x,-B_y,B_z)$:
\begin{equation}
\label{eq:gary}
\left( \begin{array}{r}
B_x \\
B_y \\
B_z \end{array} \right) =
\left( \begin{array}{rrr}
k_{31} & k_{32} & k_{33} \\
-k_{21} & -k_{22} & -k_{23} \\
k_{11} & k_{12} & k_{13} \end{array} \right)
\left( \begin{array}{r}
B_\xi \\
B_\eta \\
B_\zeta \end{array} \right),
\end{equation}
where $k_{ij}$ is a function of longitude $\phi$, latitude $\lambda$, disk center longitude $\phi_0$, $b$-angle, and $p$-angle.
\begin{equation}
\label{eq:k}
\begin{split}
k_{11} & = \cos \lambda \,[ \sin b \sin p \cos (\phi-\phi_0) + \cos p \sin(\phi-\phi_0)] - \sin \lambda \,[\cos b \sin p], \\
k_{12} & = - \cos \lambda \,[ \sin b \cos p \cos (\phi-\phi_0) - \sin p \sin(\phi-\phi_0)] + \sin \lambda \,[\cos b \cos p], \\
k_{13} & = \cos \lambda \cos b \cos (\phi-\phi_0) + \sin \lambda \sin b, \\
k_{21} & = \sin \lambda \,[ \sin b \sin p \cos (\phi-\phi_0) + \cos p \sin(\phi-\phi_0)] + \cos \lambda \,[\cos b \sin p], \\
k_{22} & = - \sin \lambda \,[ \sin b \cos p \cos (\phi-\phi_0) - \sin p \sin(\phi-\phi_0)] - \cos \lambda \,[\cos b \cos p], \\
k_{23} & = \sin \lambda \cos b \cos (\phi-\phi_0) - \cos \lambda \sin b, \\
k_{31} & = - \sin b \sin p \sin(\phi-\phi_0) + \cos p \cos(\phi-\phi_0), \\
k_{32} & = \sin b \cos p \sin(\phi-\phi_0) + \sin p \cos(\phi-\phi_0), \\
k_{33} & = - \cos b \sin (\phi-\phi_0).
\end{split}
\end{equation}

If we do not consider the finite distance correction, there is $\mu=k_{13}$. 


If we evaluate the formal uncertainties $(\sigma_{F_x}^2,\sigma_{F_y}^2,\sigma_{F_z}^2)$ in terms of $(B_x,B_y,B_z)$, we run into the problems of evaluating the covariances between $B_x$ and  $B_y$, etc., which is not straightforward. We thus rewrite Equation set~(\ref{eq:lorentz}) using (\ref{eq:gary})-(\ref{eq:k}):
\begin{equation}
\label{eq:lorentz0}
\begin{split}
F_x & = - \frac{1}{4\pi} \sum (r^{\xi \xi}_x B_\xi^2 + r^{\eta \eta}_x B_\eta^2 + r^{\zeta \zeta}_x B_\zeta^2 + r^{\xi \eta}_x B_\xi B_\eta + r^{\xi \zeta}_x B_\xi B_\zeta + r^{\eta \zeta}_x B_\eta B_\zeta) A, \\
F_y & = - \frac{1}{4\pi} \sum (r^{\xi \xi}_y B_\xi^2 + r^{\eta \eta}_y B_\eta^2 + r^{\zeta \zeta}_y B_\zeta^2 + r^{\xi \eta}_y B_\xi B_\eta + r^{\xi \zeta}_y B_\xi B_\zeta + r^{\eta \zeta}_y B_\eta B_\zeta) A, \\
F_z & = \frac{1}{8\pi} \sum (r^{\xi \xi}_z B_\xi^2 + r^{\eta \eta}_z B_\eta^2 + r^{\zeta \zeta}_z B_\zeta^2 + r^{\xi \eta}_z B_\xi B_\eta + r^{\xi \zeta}_z B_\xi B_\zeta + r^{\eta \zeta}_z B_\eta B_\zeta) A,
\end{split}
\end{equation}
where
\begin{equation}
\label{eq:r}
\left( \begin{array}{r}
r^{\xi \xi}_x \\
r^{\eta \eta}_x \\
r^{\zeta \zeta}_x \\
r^{\xi \eta}_x \\
r^{\xi \zeta}_x \\
r^{\eta \zeta}_x \\
\end{array} \right) = 
\left( \begin{array}{c}
k_{11} k_{31} \\
k_{12} k_{32} \\
k_{13} k_{33} \\
k_{12} k_{31} + k_{11} k_{32} \\
k_{11} k_{33} + k_{13} k_{31} \\
k_{12} k_{33} + k_{13} k_{32} \\
\end{array} \right), \;\;
\left( \begin{array}{r}
r^{\xi \xi}_y \\
r^{\eta \eta}_y \\
r^{\zeta \zeta}_y \\
r^{\xi \eta}_y \\
r^{\xi \zeta}_y \\
r^{\eta \zeta}_y \\
\end{array} \right) = -
\left( \begin{array}{c}
k_{11} k_{21} \\
k_{12} k_{22} \\
k_{13} k_{23} \\
k_{12} k_{21} + k_{11} k_{22} \\
k_{11} k_{23} + k_{13} k_{21} \\
k_{12} k_{23} + k_{13} k_{22} \\
\end{array} \right), \;\;
\left( \begin{array}{r}
r^{\xi \xi}_z \\
r^{\eta \eta}_z \\
r^{\zeta \zeta}_z \\
r^{\xi \eta}_z \\
r^{\xi \zeta}_z \\
r^{\eta \zeta}_z \\
\end{array} \right) =
\left( \begin{array}{c}
-k_{11}^2 + k_{21}^2 + k_{31}^2 \\
-k_{12}^2 + k_{22}^2 + k_{32}^2 \\
-k_{13}^2 + k_{23}^2 + k_{33}^2 \\
-2 k_{11} k_{12} + 2 k_{21} k_{22} + 2 k_{31} k_{32} \\
-2 k_{11} k_{13} + 2 k_{21} k_{23} + 2 k_{31} k_{33} \\
-2 k_{12} k_{13} + 2 k_{22} k_{23} + 2 k_{32} k_{33} \\
\end{array} \right).
\end{equation}

Assuming each pixel is measured individually, the formal uncertainties of the Lorentz force can be written as
\begin{equation}
\label{eq:err_f}
\begin{split}
\sigma_{F_x}^2 & = \sum \left( \left( \frac{\partial F_x}{\partial B} \right )^2 \sigma^2_{B} +  \left( \frac{\partial F_x}{\partial \gamma} \right )^2 \sigma^2_{\gamma} + \left( \frac{\partial F_x}{\partial \psi} \right )^2 \sigma^2_{\psi} + 2 \frac{\partial F_x}{\partial B} \frac{\partial F_x}{\partial \gamma} \sigma_B \sigma_\gamma \rho_{B \gamma} + 2 \frac{\partial F_x}{\partial B} \frac{\partial F_x}{\partial \psi} \sigma_B \sigma_\psi \rho_{B \psi} + 2 \frac{\partial F_x}{\partial \gamma} \frac{\partial F_x}{\partial \psi} \sigma_\gamma \sigma_\psi \rho_{\gamma \psi} \right),\\
\sigma_{F_y}^2 & = \sum \left(\left( \frac{\partial F_y}{\partial B} \right )^2 \sigma^2_{B} + \left( \frac{\partial F_y}{\partial \gamma} \right )^2 \sigma^2_{\gamma} + \left( \frac{\partial F_y}{\partial \psi} \right )^2 \sigma^2_{\psi} + 2 \frac{\partial F_y}{\partial B} \frac{\partial F_y}{\partial \gamma} \sigma_B \sigma_\gamma \rho_{B \gamma} + 2 \frac{\partial F_y}{\partial B} \frac{\partial F_y}{\partial \psi} \sigma_B \sigma_\psi \rho_{B \psi} + 2 \frac{\partial F_y}{\partial \gamma} \frac{\partial F_y}{\partial \psi} \sigma_\gamma \sigma_\psi \rho_{\gamma \psi} \right), \\
\sigma_{F_z}^2 & = \sum \left( \left( \frac{\partial F_z}{\partial B} \right )^2 \sigma^2_{B} + \left( \frac{\partial F_z}{\partial \gamma} \right )^2 \sigma^2_{\gamma} + \left( \frac{\partial F_z}{\partial \psi} \right )^2 \sigma^2_{\psi} + 2 \frac{\partial F_z}{\partial B} \frac{\partial F_z}{\partial \gamma} \sigma_B \sigma_\gamma \rho_{B \gamma} + 2 \frac{\partial F_z}{\partial B} \frac{\partial F_z}{\partial \psi} \sigma_B \sigma_\psi \rho_{B \psi} + 2 \frac{\partial F_z}{\partial \gamma} \frac{\partial F_z}{\partial \psi} \sigma_\gamma \sigma_\psi \rho_{\gamma \psi} \right),
\end{split}
\end{equation}
where the nine partial derivatives (for each pixel) are
\begin{equation}
\label{eq:pdiv_f}
\begin{split}
\frac{\partial F_x}{\partial B} = & -\frac{AB}{2 \pi} (r^{\xi \xi}_x \sin^2\gamma \sin^2\psi + r^{\eta \eta}_x \sin^2\gamma \cos^2\psi + r^{\zeta \zeta}_x \cos^2\gamma - \\
& - r^{\xi \eta}_x \sin^2\gamma \sin\psi \cos\psi - r^{\xi \zeta}_x \sin\gamma \cos\gamma \sin\psi + r^{\eta \zeta}_x \sin\gamma \cos\gamma \cos\psi),\\
\frac{\partial F_y}{\partial B} = & -\frac{AB}{2 \pi} (r^{\xi \xi}_y \sin^2\gamma \sin^2\psi + r^{\eta \eta}_y \sin^2\gamma \cos^2\psi + r^{\zeta \zeta}_y \cos^2\gamma - \\
& - r^{\xi \eta}_y \sin^2\gamma \sin\psi \cos\psi - r^{\xi \zeta}_y \sin\gamma \cos\gamma \sin\psi + r^{\eta \zeta}_y \sin\gamma \cos\gamma \cos\psi),\\
\frac{\partial F_z}{\partial B} = & \frac{AB}{4 \pi} (r^{\xi \xi}_z \sin^2\gamma \sin^2\psi + r^{\eta \eta}_z \sin^2\gamma \cos^2\psi + r^{\zeta \zeta}_z \cos^2\gamma - \\
& - r^{\xi \eta}_z \sin^2\gamma \sin\psi \cos\psi - r^{\xi \zeta}_z \sin\gamma \cos\gamma \sin\psi + r^{\eta \zeta}_z \sin\gamma \cos\gamma \cos\psi),\\
\frac{\partial F_x}{\partial \gamma} = & - \frac{AB^2} {4 \pi} (2 r^{\xi \xi}_x \sin\gamma \cos\gamma \sin^2\psi + 2 r^{\eta \eta}_x \sin\gamma \cos\gamma \cos^2\psi - 2 r^{\zeta \zeta}_x \sin\gamma \cos\gamma - \\
& - 2 r^{\xi \eta}_x \sin\gamma \cos\gamma \sin\psi \cos\psi - r^{\xi \zeta}_x (\cos^2\gamma - \sin^2\gamma)\sin\psi + r^{\eta \zeta}_x (\cos^2\gamma - \sin^2\gamma)\cos\psi,\\
\frac{\partial F_y}{\partial \gamma} = & - \frac{AB^2} {4 \pi} (2 r^{\xi \xi}_y \sin\gamma \cos\gamma \sin^2\psi + 2 r^{\eta \eta}_y \sin\gamma \cos\gamma \cos^2\psi - 2 r^{\zeta \zeta}_y \sin\gamma \cos\gamma - \\
& - 2 r^{\xi \eta}_y \sin\gamma \cos\gamma \sin\psi \cos\psi - r^{\xi \zeta}_y (\cos^2\gamma - \sin^2\gamma)\sin\psi + r^{\eta \zeta}_y (\cos^2\gamma - \sin^2\gamma)\cos\psi,\\
\frac{\partial F_z}{\partial \gamma} = & \frac{AB^2} {8 \pi} (2 r^{\xi \xi}_z \sin\gamma \cos\gamma \sin^2\psi + 2 r^{\eta \eta}_z \sin\gamma \cos\gamma \cos^2\psi - 2 r^{\zeta \zeta}_z \sin\gamma \cos\gamma - \\
& - 2 r^{\xi \eta}_z \sin\gamma \cos\gamma \sin\psi \cos\psi - r^{\xi \zeta}_z (\cos^2\gamma - \sin^2\gamma)\sin\psi + r^{\eta \zeta}_z (\cos^2\gamma - \sin^2\gamma)\cos\psi,\\
\frac{\partial F_x}{\partial \psi} = &  - \frac{AB^2} {4 \pi} (2 r^{\xi \xi}_x \sin^2\gamma \sin\psi \cos\psi - 2 r^{\eta \eta}_x \sin^2\gamma \sin\psi \cos\psi - \\
& - r^{\xi \eta}_x \sin^2\gamma (\cos^2\psi - \sin^2\psi) - r^{\xi \zeta}_x \sin\gamma \cos\gamma \cos\psi - r^{\eta \zeta}_x \sin\gamma \cos\gamma \sin\psi ),\\
\frac{\partial F_y}{\partial \psi} = &  - \frac{AB^2} {4 \pi} (2 r^{\xi \xi}_y \sin^2\gamma \sin\psi \cos\psi - 2 r^{\eta \eta}_y \sin^2\gamma \sin\psi \cos\psi - \\
& - r^{\xi \eta}_y \sin^2\gamma (\cos^2\psi - \sin^2\psi) - r^{\xi \zeta}_y \sin\gamma \cos\gamma \cos\psi - r^{\eta \zeta}_y \sin\gamma \cos\gamma \sin\psi ),\\
\frac{\partial F_z}{\partial \psi} =  & \frac{AB^2} {8 \pi} (2 r^{\xi \xi}_z \sin^2\gamma \sin\psi \cos\psi - 2 r^{\eta \eta}_z \sin^2\gamma \sin\psi \cos\psi - \\
& - r^{\xi \eta}_z \sin^2\gamma (\cos^2\psi - \sin^2\psi) - r^{\xi \zeta}_z \sin\gamma \cos\gamma \cos\psi - r^{\eta \zeta}_z \sin\gamma \cos\gamma \sin\psi ). \\
\end{split}
\end{equation}

Similarly, the relative forces are
\begin{equation}
\label{eq:epsilon0}
\begin{split}
\epsilon_x & = \frac {\sum (r^{\xi \xi}_x B_\xi^2 + r^{\eta \eta}_x B_\eta^2 + r^{\zeta \zeta}_x B_\zeta^2 + r^{\xi \eta}_x B_\xi B_\eta + r^{\xi \zeta}_x B_\xi B_\zeta + r^{\eta \zeta}_x B_\eta B_\zeta) A} { \sum B^2 A}, \\
\epsilon_y & = \frac {\sum (r^{\xi \xi}_y B_\xi^2 + r^{\eta \eta}_y B_\eta^2 + r^{\zeta \zeta}_y B_\zeta^2 + r^{\xi \eta}_y B_\xi B_\eta + r^{\xi \zeta}_y B_\xi B_\zeta + r^{\eta \zeta}_y B_\eta B_\zeta) A} { \sum B^2 A}, \\
\epsilon_z & = \frac {\sum (r^{\xi \xi}_z B_\xi^2 + r^{\eta \eta}_z B_\eta^2 + r^{\zeta \zeta}_z B_\zeta^2 + r^{\xi \eta}_z B_\xi B_\eta + r^{\xi \zeta}_z B_\xi B_\zeta + r^{\eta \zeta}_z B_\eta B_\zeta) A} { \sum B^2 A}.
\end{split}
\end{equation}
The uncertainties are
\begin{equation}
\label{eq:err_epsilon}
\begin{split}
\sigma_{\epsilon_x}^2 & = \sum \left( \left( \frac{\partial \epsilon_x}{\partial B} \right )^2 \sigma^2_{B} +  \left( \frac{\partial \epsilon_x}{\partial \gamma} \right )^2 \sigma^2_{\gamma} + \left( \frac{\partial \epsilon_x}{\partial \psi} \right )^2 \sigma^2_{\psi} + 2 \frac{\partial \epsilon_x}{\partial B} \frac{\partial \epsilon_x}{\partial \gamma} \sigma_B \sigma_\gamma \rho_{B \gamma} + 2 \frac{\partial \epsilon_x}{\partial B} \frac{\partial \epsilon_x}{\partial \psi} \sigma_B \sigma_\psi \rho_{B \psi} + 2 \frac{\partial \epsilon_x}{\partial \gamma} \frac{\partial \epsilon_x}{\partial \psi} \sigma_\gamma \sigma_\psi \rho_{\gamma \psi} \right),\\
\sigma_{\epsilon_y}^2 & = \sum \left(\left( \frac{\partial \epsilon_y}{\partial B} \right )^2 \sigma^2_{B} + \left( \frac{\partial \epsilon_y}{\partial \gamma} \right )^2 \sigma^2_{\gamma} + \left( \frac{\partial \epsilon_y}{\partial \psi} \right )^2 \sigma^2_{\psi} + 2 \frac{\partial \epsilon_y}{\partial B} \frac{\partial \epsilon_y}{\partial \gamma} \sigma_B \sigma_\gamma \rho_{B \gamma} + 2 \frac{\partial \epsilon_y}{\partial B} \frac{\partial \epsilon_y}{\partial \psi} \sigma_B \sigma_\psi \rho_{B \psi} + 2 \frac{\partial \epsilon_y}{\partial \gamma} \frac{\partial \epsilon_y}{\partial \psi} \sigma_\gamma \sigma_\psi \rho_{\gamma \psi} \right), \\
\sigma_{\epsilon_z}^2 & = \sum \left( \left( \frac{\partial \epsilon_z}{\partial B} \right )^2 \sigma^2_{B} + \left( \frac{\partial \epsilon_z}{\partial \gamma} \right )^2 \sigma^2_{\gamma} + \left( \frac{\partial \epsilon_z}{\partial \psi} \right )^2 \sigma^2_{\psi} + 2 \frac{\partial \epsilon_z}{\partial B} \frac{\partial \epsilon_z}{\partial \gamma} \sigma_B \sigma_\gamma \rho_{B \gamma} + 2 \frac{\partial \epsilon_z}{\partial B} \frac{\partial \epsilon_z}{\partial \psi} \sigma_B \sigma_\psi \rho_{B \psi} + 2 \frac{\partial \epsilon_z}{\partial \gamma} \frac{\partial \epsilon_z}{\partial \psi} \sigma_\gamma \sigma_\psi \rho_{\gamma \psi} \right).
\end{split}
\end{equation}
The partial derivatives (for each pixel), using Equation set~(\ref{eq:pdiv_f}), are
\begin{equation}
\label{eq:pdiv_epsilon}
\begin{split}
\frac{\partial \epsilon_x}{\partial B} = & -\frac{4\pi} {\sum B^2 A} \frac{\partial F_x}{\partial B} - \frac{2\epsilon_x BA}{\sum B^2 A},\\
\frac{\partial \epsilon_y}{\partial B} = & -\frac{4\pi} {\sum B^2 A} \frac{\partial F_y}{\partial B} - \frac{2\epsilon_y BA}{\sum B^2 A},\\
\frac{\partial \epsilon_z}{\partial B} = &  \frac{8\pi} {\sum B^2 A} \frac{\partial F_z}{\partial B} - \frac{2\epsilon_z BA}{\sum B^2 A},\\
\frac{\partial \epsilon_x}{\partial \gamma} = & -\frac{4\pi} {\sum B^2 A} \frac{\partial F_x}{\partial \gamma},\\
\frac{\partial \epsilon_y}{\partial \gamma} = & -\frac{4\pi} {\sum B^2 A} \frac{\partial F_y}{\partial \gamma},\\
\frac{\partial \epsilon_z}{\partial \gamma} = & \frac{8\pi} {\sum B^2 A} \frac{\partial F_z} {\partial \gamma},\\
\frac{\partial \epsilon_x}{\partial \psi} = & -\frac{4\pi} {\sum B^2 A} \frac{\partial F_x}{\partial \psi},\\
\frac{\partial \epsilon_y}{\partial \psi} = & -\frac{4\pi} {\sum B^2 A} \frac{\partial F_y}{\partial \psi},\\
\frac{\partial \epsilon_z}{\partial \psi} = &  \frac{8\pi} {\sum B^2 A} \frac{\partial F_z}{\partial \psi}. \\
\end{split}
\end{equation}


\section{Surface Integral in Spherical Coordinate\protect\footnotemark}
\label{sec:sph}

\footnotetext{Added on Jan 13, 2019. Not implemented in the pipeline.}

The derivation above assumes that the unit vectors in a Heliocentric spherical coordinate $(\hat{e}_\phi,\hat{e}_\theta,\hat{e}_r)$ is equivalent to that in a local Cartesian coordinate $(\hat{e}_x,-\hat{e}_y,\hat{e}_z)$. The assumption is generally valid for typical-sized ARs. When the field of view is large, however, the integrals must be properly evaluated in a spherical coordinate.

For simplicity, we consider only the vertical component of the Lorentz force, evaluated along the radial direction $\hat{e}_{r_0}$ of a reference point $(\phi_0,\theta_0,R_{\odot})$. To assist the derivation, we define a \textit{global} Heliocentric Cartesian coordinate, where the Sun center to the $(\lambda,\phi)=(0,0)$ line is the $X$-axis; the Sun center to the $(\lambda,\phi)=(0,90)$ line (i.e., West) is the $Y$-axis, and the $Z$-axis points solar north. Note we have $\theta=90^\circ-\lambda$. The two sets of unit vectors, $(\hat{e}_r, \hat{e}_\theta, \hat{e}_\phi)$ and $(\hat{e}_X, \hat{e}_Y, \hat{e}_Z)$ are connected by

\begin{equation}
\label{eq:unitvec}
\begin{split}
\hat{e}_\phi = & -\sin\phi \;\hat{e}_X + \cos\phi \;\hat{e}_Y,\\
\hat{e}_\theta = & \cos\theta \cos\phi \;\hat{e}_X + \cos\theta \sin\phi \;\hat{e}_Y - \sin\theta \;\hat{e}_Z,\\
\hat{e}_r = & \sin\theta \cos\phi \;\hat{e}_X + \sin\theta \sin\phi \;\hat{e}_Y + \cos\theta \;\hat{e}_Z.\\
\end{split}
\end{equation}

The reference direction $\hat{e}_{z_0} = \hat{e}_{r_0}$ can then be expressed by
\begin{equation}
\label{eq:rref}
\hat{e}_{z_0} = \sin\theta_0 \cos\phi_0 \;\hat{e}_X + \sin\theta_0 \sin\phi_0 \;\hat{e}_Y + \cos\theta_0 \;\hat{e}_Z.
\end{equation}

The total Lorentz force $F_{z_0}$ along $\hat{z}_0$ is then a linear combination of the integrands for the original $(F_x, F_y, F_z)$, i.e.,
\begin{equation}
\label{eq:fn}
F_{z_0} = -\frac{1}{4\pi} \sum c_x B_x B_z A - \frac{1}{4\pi} \sum c_x B_y B_z A + \frac{1}{8\pi} \sum c_z (B^2_x + B^2_y - B^2_z) A,
\end{equation}
where the coefficients $(c_x, c_y, c_z)$ at each pixel can be easily evaluated in the global Cartesian coordinate as
\begin{equation}
\begin{split}
\label{eq:fz}
c_x = & \; \hat{e}_x \cdot \hat{e}_{z_0} = \; \hat{e}_\phi \cdot \hat{e}_{z_0} = -\sin\phi \sin\theta_0 \cos\phi_0 + \cos\phi \sin\theta_0 \sin\phi_0\\
       = & -\sin\phi \cos\lambda_0 \cos\phi_0 + \cos\phi \cos\lambda_0 \sin\phi_0,\\
c_y = &  \; \hat{e}_y \cdot \hat{e}_{z_0} = - \hat{e}_\theta \cdot \hat{e}_{z_0} = - \cos\theta \cos\phi \sin\theta_0 \cos\phi_0 - \cos\theta \sin\phi \sin\theta_0 \sin\phi_0 + \sin\theta \cos\theta_0\\
       = & -\sin\lambda \cos\phi \cos\lambda_0 \cos\phi_0 - \sin\lambda \sin\phi \cos\lambda_0 \sin\phi_0 + \cos\lambda \sin\lambda_0,\\
c_z =  & \; \hat{e}_z \cdot \hat{e}_{z_0} = \; \hat{e}_r \cdot \hat{e}_{z_0} = \;\sin\theta \cos\phi \sin\theta_0 \cos\phi_0 + \sin\theta \sin\phi \sin\theta_0 \sin\phi_0 + \cos\theta \cos\theta_0 \\
      = & \; \cos\lambda \cos\phi \cos\lambda_0 \cos\phi_0 + \cos\lambda \sin\phi \cos\lambda_0 \sin\phi_0 + \sin\lambda \sin\lambda_0 .\\
\end{split}
\end{equation}

Following Equation sets~(\ref{eq:err_f}) and~(\ref{eq:pdiv_f}), the uncertainty can be computed as
\begin{equation}
\label{eq:err_fz0}
\sigma_{F_{z_0}}^2 = \sum \left( \left( \frac{\partial F_{z_0}}{\partial B} \right )^2 \sigma^2_{B} + \left( \frac{\partial F_{z_0}}{\partial \gamma} \right )^2 \sigma^2_{\gamma} + \left( \frac{\partial F_{z_0}}{\partial \psi} \right )^2 \sigma^2_{\psi} + 2 \frac{\partial F_{z_0}}{\partial B} \frac{\partial F_{z_0}}{\partial \gamma} \sigma_B \sigma_\gamma \rho_{B \gamma} + 2 \frac{\partial F_{z_0}}{\partial B} \frac{\partial F_{z_0}}{\partial \psi} \sigma_B \sigma_\psi \rho_{B \psi} + 2 \frac{\partial F_{z_0}}{\partial \gamma} \frac{\partial F_{z_0}}{\partial \psi} \sigma_\gamma \sigma_\psi \rho_{\gamma \psi} \right),
\end{equation}
with pixel-wise values
\begin{equation}
\label{eq:pderiv_fz0}
\begin{split}
\frac{\partial F_{z_0}}{\partial B} = & c_x \frac{\partial F_x}{\partial B} + c_y \frac{\partial F_y}{\partial B} + c_z \frac{\partial F_z}{\partial B},\\
\frac{\partial F_{z_0}}{\partial \gamma} = & c_x \frac{\partial F_x}{\partial \gamma} + c_y \frac{\partial F_y}{\partial \gamma} + c_z \frac{\partial F_z}{\partial \gamma},\\
\frac{\partial F_{z_0}}{\partial \psi} = & c_x \frac{\partial F_x}{\partial \psi} + c_y \frac{\partial F_y}{\partial \psi} + c_z \frac{\partial F_z}{\partial \psi}.\\
\end{split}
\end{equation}


\section{Including the Magnetic Filling Factor\protect\footnotemark}
\label{sec:alpha}

\footnotetext{Added on May 11, 2022. Not implemented in the pipeline.}

HMI default inversion pipeline assumes that the magnetic field is spatially uniform at the instrument resolution. The magnetic filling factor, $\alpha$, is set to unity \citep{centeno2014}. This can be viewed as a uninformative prior when the field structure is unresolved at the instrument resolution. 

A new version of the pipeline recently becomes available, which allows for a variable $\alpha \in [0,1]$ \citep{grinonmarin2021}. Below, we rewrite several equation sets (\ref{eq:lorentz0}, \ref{eq:err_f}, \ref{eq:pdiv_f}) to include the $\alpha$-related terms. These include the uncertainty ($\sigma_\alpha$) and the correlation coefficients ($\rho_{\alpha B}$, $\rho_{\alpha \gamma}$, $\rho_{\alpha \psi}$).

The Lorentz forces are now:
\begin{equation}
\label{eq:lorentz0_alpha}
\begin{split}
F_x & = - \frac{1}{4\pi} \sum (r^{\xi \xi}_x B_\xi^2 + r^{\eta \eta}_x B_\eta^2 + r^{\zeta \zeta}_x B_\zeta^2 + r^{\xi \eta}_x B_\xi B_\eta + r^{\xi \zeta}_x B_\xi B_\zeta + r^{\eta \zeta}_x B_\eta B_\zeta) \alpha A, \\
F_y & = - \frac{1}{4\pi} \sum (r^{\xi \xi}_y B_\xi^2 + r^{\eta \eta}_y B_\eta^2 + r^{\zeta \zeta}_y B_\zeta^2 + r^{\xi \eta}_y B_\xi B_\eta + r^{\xi \zeta}_y B_\xi B_\zeta + r^{\eta \zeta}_y B_\eta B_\zeta) \alpha A, \\
F_z & = \frac{1}{8\pi} \sum (r^{\xi \xi}_z B_\xi^2 + r^{\eta \eta}_z B_\eta^2 + r^{\zeta \zeta}_z B_\zeta^2 + r^{\xi \eta}_z B_\xi B_\eta + r^{\xi \zeta}_z B_\xi B_\zeta + r^{\eta \zeta}_z B_\eta B_\zeta) \alpha A,
\end{split}
\end{equation}

Their uncertainties are now:
\begin{equation}
\label{eq:err_f_alpha}
\begin{split}
\sigma_{F_x}^2  = \sum & \left( \frac{\partial F_x}{\partial B} \right )^2 \sigma^2_{B} +  \left( \frac{\partial F_x}{\partial \gamma} \right )^2 \sigma^2_{\gamma} + \left( \frac{\partial F_x}{\partial \psi} \right )^2 \sigma^2_{\psi} + \left( \frac{\partial F_x}{\partial \alpha} \right )^2 \sigma^2_{\alpha} + 2 \frac{\partial F_x}{\partial B} \frac{\partial F_x}{\partial \gamma} \sigma_B \sigma_\gamma \rho_{B \gamma} + 2 \frac{\partial F_x}{\partial B} \frac{\partial F_x}{\partial \psi} \sigma_B \sigma_\psi \rho_{B \psi} + \\
	& + 2 \frac{\partial F_x}{\partial \gamma} \frac{\partial F_x}{\partial \psi} \sigma_\gamma \sigma_\psi \rho_{\gamma \psi} + 2 \frac{\partial F_x}{\partial \alpha} \frac{\partial F_x}{\partial B} \sigma_\alpha \sigma_B  \rho_{\alpha B} + 2 \frac{\partial F_x}{\partial \alpha} \frac{\partial F_x}{\partial \gamma} \sigma_\alpha \sigma_\gamma \rho_{\alpha \gamma} + 2 \frac{\partial F_x}{\partial \alpha} \frac{\partial F_x}{\partial \psi} \sigma_\alpha \sigma_\psi \rho_{B \psi},\\
\sigma_{F_y}^2 = \sum & \left( \frac{\partial F_y}{\partial B} \right )^2 \sigma^2_{B} + \left( \frac{\partial F_y}{\partial \gamma} \right )^2 \sigma^2_{\gamma} + \left( \frac{\partial F_y}{\partial \psi} \right )^2 \sigma^2_{\psi} + \left( \frac{\partial F_y}{\partial \alpha} \right )^2 \sigma^2_{\alpha} + 2 \frac{\partial F_y}{\partial B} \frac{\partial F_y}{\partial \gamma} \sigma_B \sigma_\gamma \rho_{B \gamma} + 2 \frac{\partial F_y}{\partial B} \frac{\partial F_y}{\partial \psi} \sigma_B \sigma_\psi \rho_{B \psi} + \\
	& + 2 \frac{\partial F_y}{\partial \gamma} \frac{\partial F_y}{\partial \psi} \sigma_\gamma \sigma_\psi \rho_{\gamma \psi} + 2 \frac{\partial F_y}{\partial \alpha} \frac{\partial F_y}{\partial B} \sigma_\alpha \sigma_B  \rho_{\alpha B} + 2 \frac{\partial F_y}{\partial \alpha} \frac{\partial F_y}{\partial \gamma} \sigma_\alpha \sigma_\gamma \rho_{\alpha \gamma} + 2 \frac{\partial F_y}{\partial \alpha} \frac{\partial F_y}{\partial \psi} \sigma_\alpha \sigma_\psi \rho_{B \psi},\\
\sigma_{F_z}^2 = \sum & \left( \frac{\partial F_z}{\partial B} \right )^2 \sigma^2_{B} + \left( \frac{\partial F_z}{\partial \gamma} \right )^2 \sigma^2_{\gamma} + \left( \frac{\partial F_z}{\partial \psi} \right )^2 \sigma^2_{\psi} + \left( \frac{\partial F_z}{\partial \alpha} \right )^2 \sigma^2_{\alpha} + 2 \frac{\partial F_z}{\partial B} \frac{\partial F_z}{\partial \gamma} \sigma_B \sigma_\gamma \rho_{B \gamma} + 2 \frac{\partial F_z}{\partial B} \frac{\partial F_z}{\partial \psi} \sigma_B \sigma_\psi \rho_{B \psi} + \\
	& + 2 \frac{\partial F_z}{\partial \gamma} \frac{\partial F_z}{\partial \psi} \sigma_\gamma \sigma_\psi \rho_{\gamma \psi} + 2 \frac{\partial F_z}{\partial \alpha} \frac{\partial F_z}{\partial B} \sigma_\alpha \sigma_B  \rho_{\alpha B} + 2 \frac{\partial F_z}{\partial \alpha} \frac{\partial F_z}{\partial \gamma} \sigma_\alpha \sigma_\gamma \rho_{\alpha \gamma} + 2 \frac{\partial F_z}{\partial \alpha} \frac{\partial F_z}{\partial \psi} \sigma_\alpha \sigma_\psi \rho_{B \psi}.\\
\end{split}
\end{equation}

The twelve partial derivatives (per pixel) are now:
\begin{equation}
\label{eq:pdiv_f_alpha}
\begin{split}
\frac{\partial F_x}{\partial B} = & -\frac{\alpha AB}{2 \pi} (r^{\xi \xi}_x \sin^2\gamma \sin^2\psi + r^{\eta \eta}_x \sin^2\gamma \cos^2\psi + r^{\zeta \zeta}_x \cos^2\gamma - \\
& - r^{\xi \eta}_x \sin^2\gamma \sin\psi \cos\psi - r^{\xi \zeta}_x \sin\gamma \cos\gamma \sin\psi + r^{\eta \zeta}_x \sin\gamma \cos\gamma \cos\psi),\\
\frac{\partial F_y}{\partial B} = & -\frac{\alpha AB}{2 \pi} (r^{\xi \xi}_y \sin^2\gamma \sin^2\psi + r^{\eta \eta}_y \sin^2\gamma \cos^2\psi + r^{\zeta \zeta}_y \cos^2\gamma - \\
& - r^{\xi \eta}_y \sin^2\gamma \sin\psi \cos\psi - r^{\xi \zeta}_y \sin\gamma \cos\gamma \sin\psi + r^{\eta \zeta}_y \sin\gamma \cos\gamma \cos\psi),\\
\frac{\partial F_z}{\partial B} = & \frac{\alpha AB}{4 \pi} (r^{\xi \xi}_z \sin^2\gamma \sin^2\psi + r^{\eta \eta}_z \sin^2\gamma \cos^2\psi + r^{\zeta \zeta}_z \cos^2\gamma - \\
& - r^{\xi \eta}_z \sin^2\gamma \sin\psi \cos\psi - r^{\xi \zeta}_z \sin\gamma \cos\gamma \sin\psi + r^{\eta \zeta}_z \sin\gamma \cos\gamma \cos\psi),\\
\frac{\partial F_x}{\partial \gamma} = & - \frac{\alpha AB^2} {4 \pi} (2 r^{\xi \xi}_x \sin\gamma \cos\gamma \sin^2\psi + 2 r^{\eta \eta}_x \sin\gamma \cos\gamma \cos^2\psi - 2 r^{\zeta \zeta}_x \sin\gamma \cos\gamma - \\
& - 2 r^{\xi \eta}_x \sin\gamma \cos\gamma \sin\psi \cos\psi - r^{\xi \zeta}_x (\cos^2\gamma - \sin^2\gamma)\sin\psi + r^{\eta \zeta}_x (\cos^2\gamma - \sin^2\gamma)\cos\psi,\\
\frac{\partial F_y}{\partial \gamma} = & - \frac{\alpha AB^2} {4 \pi} (2 r^{\xi \xi}_y \sin\gamma \cos\gamma \sin^2\psi + 2 r^{\eta \eta}_y \sin\gamma \cos\gamma \cos^2\psi - 2 r^{\zeta \zeta}_y \sin\gamma \cos\gamma - \\
& - 2 r^{\xi \eta}_y \sin\gamma \cos\gamma \sin\psi \cos\psi - r^{\xi \zeta}_y (\cos^2\gamma - \sin^2\gamma)\sin\psi + r^{\eta \zeta}_y (\cos^2\gamma - \sin^2\gamma)\cos\psi,\\
\frac{\partial F_z}{\partial \gamma} = & \frac{\alpha AB^2} {8 \pi} (2 r^{\xi \xi}_z \sin\gamma \cos\gamma \sin^2\psi + 2 r^{\eta \eta}_z \sin\gamma \cos\gamma \cos^2\psi - 2 r^{\zeta \zeta}_z \sin\gamma \cos\gamma - \\
& - 2 r^{\xi \eta}_z \sin\gamma \cos\gamma \sin\psi \cos\psi - r^{\xi \zeta}_z (\cos^2\gamma - \sin^2\gamma)\sin\psi + r^{\eta \zeta}_z (\cos^2\gamma - \sin^2\gamma)\cos\psi,\\
\frac{\partial F_x}{\partial \psi} = &  - \frac{\alpha AB^2} {4 \pi} (2 r^{\xi \xi}_x \sin^2\gamma \sin\psi \cos\psi - 2 r^{\eta \eta}_x \sin^2\gamma \sin\psi \cos\psi - \\
& - r^{\xi \eta}_x \sin^2\gamma (\cos^2\psi - \sin^2\psi) - r^{\xi \zeta}_x \sin\gamma \cos\gamma \cos\psi - r^{\eta \zeta}_x \sin\gamma \cos\gamma \sin\psi ),\\
\frac{\partial F_y}{\partial \psi} = &  - \frac{\alpha AB^2} {4 \pi} (2 r^{\xi \xi}_y \sin^2\gamma \sin\psi \cos\psi - 2 r^{\eta \eta}_y \sin^2\gamma \sin\psi \cos\psi - \\
& - r^{\xi \eta}_y \sin^2\gamma (\cos^2\psi - \sin^2\psi) - r^{\xi \zeta}_y \sin\gamma \cos\gamma \cos\psi - r^{\eta \zeta}_y \sin\gamma \cos\gamma \sin\psi ),\\
\frac{\partial F_z}{\partial \psi} =  & \frac{\alpha AB^2} {8 \pi} (2 r^{\xi \xi}_z \sin^2\gamma \sin\psi \cos\psi - 2 r^{\eta \eta}_z \sin^2\gamma \sin\psi \cos\psi - \\
& - r^{\xi \eta}_z \sin^2\gamma (\cos^2\psi - \sin^2\psi) - r^{\xi \zeta}_z \sin\gamma \cos\gamma \cos\psi - r^{\eta \zeta}_z \sin\gamma \cos\gamma \sin\psi ), \\
\frac{\partial F_x}{\partial \alpha} = & - \frac{A}{4\pi} (r^{\xi \xi}_x B_\xi^2 + r^{\eta \eta}_x B_\eta^2 + r^{\zeta \zeta}_x B_\zeta^2 + r^{\xi \eta}_x B_\xi B_\eta + r^{\xi \zeta}_x B_\xi B_\zeta + r^{\eta \zeta}_x B_\eta B_\zeta),\\
\frac{\partial F_y}{\partial \alpha} = & - \frac{A}{4\pi} (r^{\xi \xi}_y B_\xi^2 + r^{\eta \eta}_y B_\eta^2 + r^{\zeta \zeta}_y B_\zeta^2 + r^{\xi \eta}_y B_\xi B_\eta + r^{\xi \zeta}_y B_\xi B_\zeta + r^{\eta \zeta}_y B_\eta B_\zeta), \\
\frac{\partial F_z}{\partial \alpha} = & \frac{A}{8\pi} (r^{\xi \xi}_z B_\xi^2 + r^{\eta \eta}_z B_\eta^2 + r^{\zeta \zeta}_z B_\zeta^2 + r^{\xi \eta}_z B_\xi B_\eta + r^{\xi \zeta}_z B_\xi B_\zeta + r^{\eta \zeta}_z B_\eta B_\zeta). \\
\end{split}
\end{equation}

In spherical coordinate, the Lorentz force along $\hat{z}_0$ is
\begin{equation}
\label{eq:fn_alpha}
F_{z_0} = -\frac{1}{4\pi} \sum c_x B_x B_z \alpha A - \frac{1}{4\pi} \sum c_x B_y B_z \alpha A + \frac{1}{8\pi} \sum c_z (B^2_x + B^2_y - B^2_z) \alpha A.
\end{equation}

The corresponding uncertainty is
\begin{equation}
\label{eq:err_fz0_alpha}
\begin{split}
\sigma_{F_{z_0}}^2 = \sum & \left( \frac{\partial F_{z_0}}{\partial B} \right )^2 \sigma^2_{B} + \left( \frac{\partial F_{z_0}}{\partial \gamma} \right )^2 \sigma^2_{\gamma} + \left( \frac{\partial F_{z_0}}{\partial \psi} \right )^2 \sigma^2_{\psi} + \left( \frac{\partial F_{z_0}}{\partial \alpha} \right )^2 \sigma^2_{\alpha} + \\
& + 2 \frac{\partial F_{z_0}}{\partial B} \frac{\partial F_{z_0}}{\partial \gamma} \sigma_B \sigma_\gamma \rho_{B \gamma} + 2 \frac{\partial F_{z_0}}{\partial B} \frac{\partial F_{z_0}}{\partial \psi} \sigma_B \sigma_\psi \rho_{B \psi} + 2 \frac{\partial F_{z_0}}{\partial \gamma} \frac{\partial F_{z_0}}{\partial \psi} \sigma_\gamma \sigma_\psi \rho_{\gamma \psi} + \\
& + 2 \frac{\partial F_{z_0}}{\partial \alpha} \frac{\partial F_{z_0}}{\partial B}  \sigma_\alpha \sigma_B  \rho_{\alpha B} + 2 \frac{\partial F_{z_0}}{\partial \alpha} \frac{\partial F_{z_0}}{\partial \gamma} \sigma_\alpha \sigma_\gamma \rho_{\alpha \gamma} + 2 \frac{\partial F_{z_0}}{\partial \alpha} \frac{\partial F_{z_0}}{\partial \psi} \sigma_\alpha \sigma_\psi \rho_{\alpha \psi},\\
\end{split}
\end{equation}

with pixel-wise values
\begin{equation}
\label{eq:pderiv_fz0}
\begin{split}
\frac{\partial F_{z_0}}{\partial B} = & c_x \frac{\partial F_x}{\partial B} + c_y \frac{\partial F_y}{\partial B} + c_z \frac{\partial F_z}{\partial B},\\
\frac{\partial F_{z_0}}{\partial \gamma} = & c_x \frac{\partial F_x}{\partial \gamma} + c_y \frac{\partial F_y}{\partial \gamma} + c_z \frac{\partial F_z}{\partial \gamma},\\
\frac{\partial F_{z_0}}{\partial \psi} = & c_x \frac{\partial F_x}{\partial \psi} + c_y \frac{\partial F_y}{\partial \psi} + c_z \frac{\partial F_z}{\partial \psi},\\
\frac{\partial F_{z_0}}{\partial \alpha} = & c_x \frac{\partial F_x}{\partial \alpha} + c_y \frac{\partial F_y}{\partial \alpha} + c_z \frac{\partial F_z}{\partial \alpha}.\\
\end{split}
\end{equation}

%
%

\end{document}